\documentclass[%
onecolumn,
preprint, 
bibnotes,
amsmath,amssymb,
aps,
pra,
floatfix,
]{revtex4-2}
\usepackage{url,hyperref,microtype}
\usepackage {xcolor}

\usepackage{amsmath}
\usepackage{amssymb}
\usepackage{amsfonts}
\usepackage{mathtools}
\usepackage{bbm}
\usepackage{mathrsfs}
\usepackage{stmaryrd}

\usepackage{dsfont}
\usepackage{units}

\DeclareMathOperator{\diag}{diag}

\newcommand{\1}{\mathbf{1}}

\newcommand{\wegThor}[1]{\vec{#1}}
\newcommand{\VA}{\wegThor{A}}

\newcommand{\Vu}{\wegThor{u}}
\newcommand{\VN}{\wegThor{N}}
\newcommand{\VU}{\wegThor{U}}

\newcommand{\Wr}{\overline{W}}

\newcommand{\Eq}[1]{Eq.~(\ref{#1})}

\newcommand{\Fig}[1]{Fig.~\ref{#1}}
\newcommand{\Tab}[1]{Table~\ref{#1}}

\begin{document}
\title[Routing by spontaneous synchronization]{Routing by spontaneous synchronization} 
\author{Maik Schünemann${}^1$, and Udo Ernst${}^1$}

\address{
    ${}^1$ Computational Neurophysics Lab, Institute for Theoretical Physics, University of Bremen, Bremen, Germany}
\date{\today}

\begin{abstract}
    Selective attention allows to process stimuli which are
    behaviorally relevant, while attenuating distracting information.
    However, it is an open question what mechanisms implement
    selective routing, and how they are engaged in dependence
    on behavioral need.
    Here we introduce a novel framework for selective processing
    by spontaneous synchronization. Input signals become organized into
    'avalanches' of synchronized spikes which propagate to target populations.
    Selective attention enhances spontaneous synchronization and
    boosts signal transfer by a simple disinhibition of a control
    population, without requiring changes in synaptic weights.
    Our framework is fully analytically tractable and provides a complete
    understanding of all stages of the routing mechanism, yielding
    closed-form expressions for input-output correlations.
    Interestingly, although gamma oscillations can naturally occur
    through a recurrent dynamics, we can formally show that
    the routing mechanism itself does not require such oscillatory
    activity and works equally well if synchronous events would be
    randomly shuffled over time.
    Our framework explains a large range of physiological findings
    in a unified framework and makes specific predictions
    about putative control mechanisms
    and their effects on neural dynamics.
\end{abstract}

\maketitle

\vfill\pagebreak

\vfill\pagebreak

\section{Introduction}
At every moment, the brain processes enormous amounts of sensory signals with limited computational resources. Feed forward connections in the visual cortex show a pattern of convergence leading to larger receptive fields and more specialized responses in further downstream areas. The convergence of downstream connections thus requires the brain to flexibly route the currently behaviorally relevant information for further processing, while ignoring irrelevant information. A prominent example of a brain function requiring flexible signal routing is selective visual attention.

In their seminal study, Moran and Desimone \citep{moran1985selective} observed that when an attended and a non-attended stimulus were present in the receptive field of a neuron in visual area V4, the neuron behaves as if only the attended stimulus was present. Selective processing appears to be a robust mechanism which extends easily to more general situations such as selectively routing arbitrary time-varying signals originating from complex visual shapes\citep{grothe2018attention}. These results imply a remarkable flexibility of information processing in the brain to establish efficient communication channels between sets of neural populations, and to switch these based on attentional focus. 

Interestingly, selective routing appears to be established by changing effective interactions between the neural populations sending the visual signals (such as neurons in areas V1/V2) and populations in V4 receiving these signals, rather than enhancing the stimulus representation in the sender: In comparison with the strong gating effects reaching factors to two and more \citep{grothe2018attention}, attentional firing rate modulations observed in V1 are typically reported to be only around +20\% -- +30\%~\citep{roelfsema1998object,thiele2009additive,chen2008task}, while the representation of the visual signal is even slightly \emph{attenuated} by about 10\% \citep{grothe2018attention}. These differences of effect sizes for rate modulations and selective signal routing suggest a nonlinear mechanism amplifying selectivity of information routing.

Since changing synaptic efficacies on the fast timescales involved is biophysically unplausible, investigations therefore focused on putative collective phenomena suitable to rapidly modulate effective interactions. Putative routing mechanism between cortical populations are reviewed in~\citep{kohn2020principles}, but one idea received particular attention since it is well-supported by experimental evidence: The Communication-through-Coherence \citep{fries2005mechanism,fries2015rhythms} or Routing-by-synchrony~\citep{kreiter2020synchrony} framework posits that neural populations in V1 and V4 are in a strong oscillatory regime and establish communication channels by establishing favorable phase relationships between the populations in the pathway of the attended signals while pushing the population processing the non-attended stimulus in a non-favorable, or random phase relationship with the receiving population. Several models use this mechanism to explain selective signal routing, for example \cite{mishra2006selective,zeitler2008biased,masuda2009selective,montijn2012divisive,tiesinga2010mechanisms,harnack2015model}. In parallel, experimental studies established that directing attention to one stimulus in the receptive field of a V1 populations increases its $\gamma$-band power, increases the phase coherence between this population and the receiving V4 population, and the synchrony of $\gamma$ oscillations in V4 \citep{taylor2005coherent,fries2001modulation,fries2008effects,grothe2012switching,bosman2012attentional,womelsdorf2007modulation,grothe2018attention}.


However, having to establish favorable phase relationships immediately raises the question how such a fine temporal coordination \emph{between} receiving and transmitting populations is organized. Although schemes for establishing the 'correct' phases have been proposed (e.g.~\cite{harnack2015model}), it is an open question if selective processing compatible with physiological data can also be achieved by simpler mechanisms.

An alternative possibility to coordinate communication between neural populations could be to modulate the level of synchronous activity within the sending population \citep{kumar2010spiking,brette2012computing}, and to detect enhanced coincidences of incoming spikes in the receiving population. For example, a sending population operating in an asynchronous irregular dynamical regime would have low levels of population synchrony and approximate a Poissonian spike statistics, whereas high levels of synchrony would be exhibited by sending populations operating near a critical point where spiking activity features power-law distributed neural avalanches~\citep{eurich2002finite,beggs2003neuronal,beggs2008criticality,yu2014scale,priesemann2013neuronal}. On the receiving side, pyramidal neurons which in vivo show a robust synchronicity gain could act as coincidence detectors and respond more strongly to a synchronous volley of arriving action potentials compared to the same number of spikes distributed randomly over a longer time interval \citep{konig1996integrator,prescott2006nonlinear,mejias2008role,AZOUZ2003513,borgers2014approximate,haider2007enhancement,ratte2013impact,fontaine2014spike}. In the brain, such synchronicity gain is seen robustly for different levels of the presynaptic mean firing rate, and not just in the vicinity of the firing threshold as in the widely used integrate and fire neuron models \citep{huang2016adaptive,mensi2016enhanced}.

The fundamental feasibility of such schemes for routing \citep{Palmigiano2017} and feature integration \citep{tomen2014marginally,tomen2019role} has been demonstrated, but their realization of enhanced synchrony as a mechanism for attentional modulation bases on increasing recurrent connection strengths. Given that switches of attention can be completed within time spans of $\unit[100]{ms}$ to $\unit[300]{ms}$~\citep{carlson2006speed}, such sudden changes in interaction strengths are hard to realize in a biophysically plausible manner. In addition, the systems considered so far are eluding a thorough mathematical analysis, which makes it difficult to identify suitable working regimes for selective processing, and to analyze how selective processing depends on network structure and its dynamics. Our intention with this study is to close this gap, and to establish a novel, biophysically realistic and mathematically tractable framework for selective processing in the brain.

In this contribution we propose that \emph{enhancement of the population synchrony} in the populations processing the attended stimulus can establish a selective routing channel to the receiving population. We devise a physiologically plausible control mechanism in which attention disinhibits a small control population which enhances the population synchrony of the transmitting population processing the attended signal. We show that our model jointly reproduces a large number of key physiological signatures of selective information routing. Most importantly, we use a novel, mathematically tractable framework~\citep{schunemann2022rigorous} which allows to investigate the routing mechanism of the model analytically and to identify optimal routing configurations in dependence of model parameters and dynamical regimes of the populations in the network.

\vfill\pagebreak
\section{Results}

\subsection{Neural populations}\label{sec:np}

The basic building block for our study is a homogeneously coupled population of $N$ non-leaky integrate-and-fire neurons. Each neuron is described by its (normalized) membrane potential $u$ with $u=0$ being the resting state, and $u_{thr}=1$ its firing threshold. External excitatory input drives the membrane potential towards firing threshold. If a neuron fires, its membrane potential is reduced by $u_{thr}$ (reset), while the membrane potentials of all other neurons in the population are instantly increased by an amount $\alpha/N$. This recurrent input can cause other neurons to become active and thus lead to an avalanche propagating through the population. Throughout our study, we assume that the recurrent dynamics acts fast compared to the external drive (time scale separation) which simplifies mathematical analysis.

\begin{figure}[ht]
     \centering
    \includegraphics[width=\textwidth]{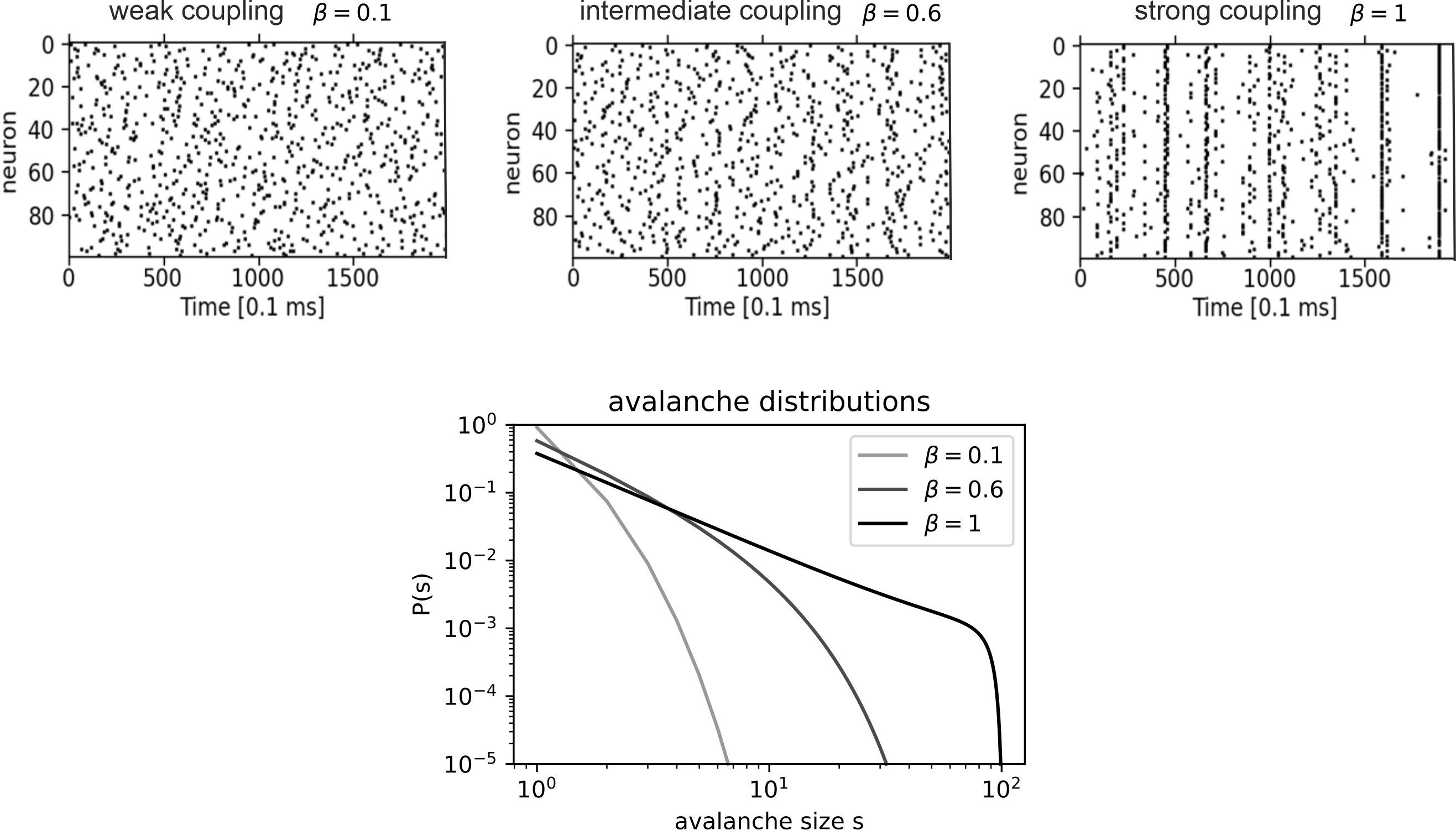}
    \caption{Modulations of population synchrony by recurrent connection strength in the Eurich-Herrmann-Ernst model \citep{eurich2002finite} Top row: Raster plot of $n=100$ model neurons, which are coupled all-to-all with homogeneous coupling weight $\beta \alpha(n)$ for weak, intermediate and strong/critical coupling strenghts. Bottom row: Corresponding analytical avalanche size distributions. For $\beta=1$, the avalanche distribution displays a power-law with an exponent of $-1.5$.}
    \label{fig:ehe-rasters-avalanches}
\end{figure}

We choose this model since it has been successfully used to model neural avalanche formation \citep{eurich2002finite} with a similar dynamics as in biological networks \citep{beggs2003neuronal}. When driven by a stochastic external input, the population generates synchronous events (avalanches) whose sizes increase with recurrent coupling strength. 
Figure~\ref{fig:ehe-rasters-avalanches} exemplifies the transition from a weak coupling with largely unsynchronized activity to a critical regime where avalanches occur over all length scales, exhibiting a power law for the avalanche size distribution. 

Recently, substantial progress has been made in extending the mathematical analysis of this model to arbitrary (positive) coupling matrices \citep{schunemann2022rigorous}. The extension allows application of the framework also to \emph{structured systems} consisting of multiple populations as e.g. in the visual system. This opportunity lead us to compose a routing model from the building blocks introduced here.

\subsection{Routing network}


The routing model is set up to reproduce the experimentally well investigated situation \citep{taylor2005coherent,grothe2012switching,bosman2012attentional,grothe2018attention} in which two stimuli lie in the receptive field of one population in area V4, but are processed by distinct populations in visual area V1. This situation is modelled by two V1 populations $A,B$ which are driven by distinct input signals, and which send their output signals to one receiving V4 population $C$. Dynamics and internal coupling of each population is as described above in section~\ref{sec:np}. Connections between populations are feed-forward and also all-to-all. In order to model the synchronicity gain of pyramidal neurons in V4 and to precisely control its strength while keeping its implementation simple to allow mathematical treatment, we introduce a threshold $\theta$ and only pass avalanches with a size larger than this threshold forward to the $V4$ population $C$. 

For quantifying the effectiveness of signal routing, visual stimuli were tagged in previous studies~\citep{grothe2018attention,harnack2015model} by independent random modulations of their luminance ('flicker'). Computing correlation measures between flicker and neural responses then allowed to determine performance and selectivity of signal routing. We adopt this approach for our model and generate input signals arising from the two stimuli with a mean intensity modulated by independent realisations $f_A,f_B$ of a 'flicker' signal changing its values every $10\unit{ms}$ to a randomly chosen level between $-1$ and $1$.

\begin{figure}[ht]
    \centering
    \includegraphics[width=\textwidth]{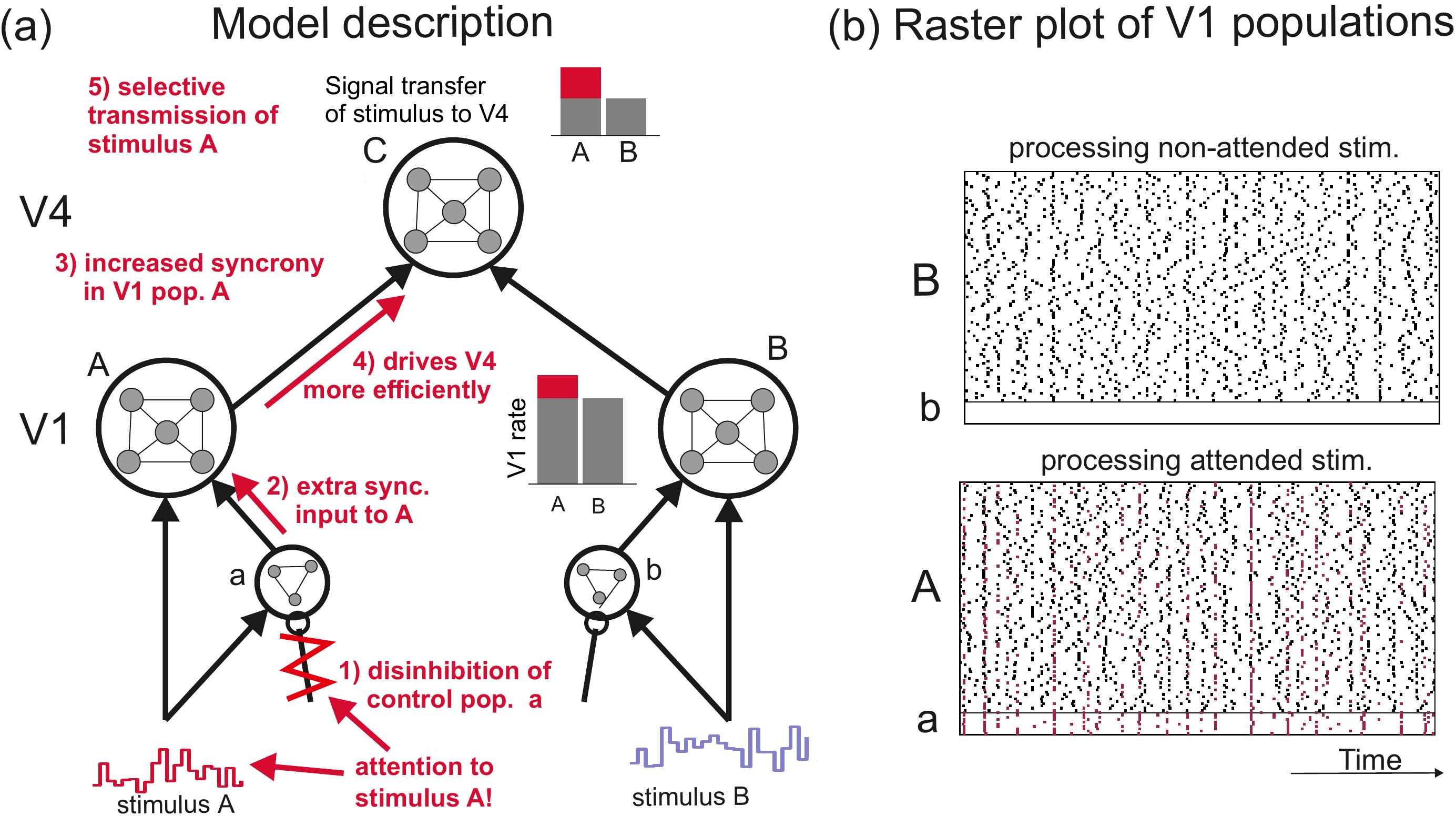}
    \caption{Model setup, routing mechanism, and spiking dynamics.
    Panel (a): The model consists of two V1 populations $A,B$ and one V4 population $C$, which are realized as  all-to-all coupled (homogeneous) networks of $n$ neurons, while the smaller 'control' populations $a,b$ consists of $m<n$ units. Connections between population are feedforward in direction of the arrows. Each neuron in the sending population is coupled to all neurons in the receiving population. The time varying stimuli consists of the instantaneous intensity of the input to the corresponding control and V1 populations. 
    Red arrows and text illustrate the mechanism of preferential signal routing by attention (see detailed explanations in text).
    Panel (b): Raster plots of spiking activity in the V1 populations processing the attended and non-attended stimulus, and spiking activity in the control populations for parameters $N=100,\beta=0.76$.
    Spikes occurring in avalanches started in V1 are illustrated by black squares, while spikes occurring in avalanches which started in the control population are illustrated by red square dots.} 
    \label{fig:routing-illus}
    
\end{figure}

Without attention, there is no advantage to either stimulus since the sending populations $A$ and $B$ drive the receiving population on average to the same extent. For establishing selective processing of $A$ or $B$, we now hypothesize that it suffices to enhance synchronization in the V1 population processing the signal of the attended stimulus. Because selective attention can be rapidly switched between different targets~\citep{carlson2006speed}, we exclude the possibility to model such enhancement by increasing the recurrent weights, since changes in synaptic efficacy typically take more time to manifest. Instead, we assume that the V1 populations $A$ and $B$ are accompanied by smaller 'control populations' $a$ and $b$, respectively. Control populations are also driven by the corresponding external input signals and provide extra input to their corresponding V1 population, see figure~\ref{fig:routing-illus}. They operate closer to the critical regime than their target populations $A$ and $B$, i.e. they possess a larger recurrent coupling strength. 

One of the two stimuli -- here without loss of generality stimulus A -- is behaviorally relevant and is attended. In our model, the control populations are inhibited (silenced) if attention is not directed to the corresponding stimulus. Directing attention to the input signal of one V1 population disinhibits and thereby activates the corresponding control population. The resulting mechanism of selective signal routing is illustrated in figure~\ref{fig:routing-illus}: Disinhibition of the critical control population $a$ adds extra input with a high level of synchronicity to V1 population $A$. Such extra input not only increases the firing rate of $A$ compared to population $B$, but also the level of synchrony in the population activity of $A$ (raster plots in Fig~\ref{fig:routing-illus}). The combined increase of firing rate and synchrony leads to a supralinear increase in the effective drive to V4 population $C$, since it responds more strongly to synchronous input. 

In the following analysis, we will show that these effects in combination establish a preferential communication channel between the attended stimulus and population $C$. 




\subsection{Analytical treatment of model dynamics}\label{sec:analytical-treatment}


After introducing the routing model and demonstrating its ability to selectively transfer stimulus signals to the receiving V4 population, we want to investigate how spontaneous synchronization supports this function. For this purpose, we can use the mathematical theory developed in~\citep{schunemann2022rigorous} to better understand network dynamics and signal transfer by computing avalanche size distributions and signal correlations between the different model stages.

\subsubsection{Avalanche size distributions}


We start by evaluating avalanche size distributions. Since at a single time step, external input is given to just one neuron, avalanches can either start in population $a$, $A$, or $B$ (Fig~\ref{fig:routing-illus}). Conditioned on the event that an avalanche was started by external input, one obtains the following distributions for the three distinct cases:
\begin{itemize}
\item
    $P_B(n_B, n_C)$ is the probability that an avalanche triggered by external input to V1 population $B$ consists of $n_B$ units in $B$ and activates $n_C$ units in V4 population $C$. Here, $n_B > 1$.
\item
    $P_A(n_A, n_C)$ is the probability that an avalanche triggered by external input to V1 population $A$ consists of $n_A$ units in $A$ and activates $n_C$ units in V4 population $C$. Here, $n_A > 1$.
\item 
    $P_a(m_a, n_A, n_C)$ is the probability that an avalanche triggered by external input to control population $a$ consists of $m_a$ units in $a$, then activates $n_A$ units in V1 population $A$, and finally activates $n_C$ units in V4 population $C$.
    Here, $m_a > 1$.
\end{itemize}
For these probability distributions, the theory developed in~\citep{schunemann2022rigorous} provides exact expressions involving the coupling matrix $W$, which are explicitly stated in the Methods section (Eqs.~\ref{eq:PV1B},~\ref{eq:PV1A},~\ref{eq:PC})



The resulting distributions do not take the synchronicity threshold $\theta$ into account, which implements the condition that avalanches in V4 population $C$ are triggered only if at least $\theta$ presynaptic units become activated. With $\theta$ being active, the distributions $P_B, P_A, P_a$ become $P^\theta_B,P^\theta_A,P^\theta_a$, respectively. Here the idea is to collapse all cases where the number of activated presynaptic units does not reach the threshold $\theta$ into the probability $P^\theta$ with $n_C=0$:

\begin{align}\label{eq:Pthetas}
    P^{\theta}_{B}(n_B, n_C) &=
    \begin{cases}
        \sum_{n'_C=0}^{n} P_{B}(n_B, n'_C) & \mbox{if } n_B<\theta \text{ and } n_C = 0\\
        0 & \mbox{if } n_B < \theta \text{ and } n_C>0 \\
        P_B(n_B, n_C) & \mbox{if } n_B \geq \theta
    \end{cases} \\
    P^{\theta}_A(n_A,n_C) & = P^{\theta}_B(n_A,n_C) \\
    P^{\theta}_{a}(n_a,n_A,n_C) &= 
    \begin{cases}
        \sum_{n'_C=0}^n P_{a}(m_a, n_A, n'_C) & \mbox{if } n_A<\theta \text{ and } n_C=0\\
        0 & \mbox{if } n_A<\theta \text{ and } n_C>0 \\
        P_a(m_a, n_A, n_C) & \mbox{if } n_A \geq \theta 
    \end{cases}\label{eq:Pthetaa}
\end{align}



The avalanche distributions $P^{\theta}_B(n_B), P^{\theta}_A(n_A), P^{\theta}_a(m_a)$, and $P^{\theta}_C(n_C)$ observed in the separate subnetworks (populations) are given by a mixture of the marginals of the avalanche distributions in Eqs.~\ref{eq:Pthetas}. For $B$ and $a$, the corresponding expressions $P^{\theta}_B(n_B)$ and $P^{\theta}_a(m_a)$ are very simple (see Methods, Eqs.~\ref{eq:avsV1B},~\ref{eq:avsC}) since avalanches in these two populations could only have originated by direct external input. For $P^{\theta}_A(n_A)$, avalanches could have been triggered either directly by external input, or by input from control population $a$, so both contributions have to be weighted by the corresponding probabilities that external input triggered either $A$ or $a$ (see Methods, Eq.~\ref{eq:avsV1A}). Finally, avalanches in $C$ could have been triggered by external input to either $A$, $B$, or $a$, so $P^{\theta}_C(n_C)$ is a weighted mixture of three marginals (see Methods, Eq.\ref{eq:avsV4})

\subsubsection{Signal correlations}

%

For assessing signal correlations between different neural populations, or between stimulus signals and neural responses, we compute Pearson correlation coefficients via
\begin{equation}
c(X,Y) = \frac{\operatorname{cov}(X,Y)}{\sqrt{\operatorname{var}(X)\operatorname{var}(Y)}} \text{ .}
\end{equation}
Activities in neural populations $\{X, Y\} \in \{A,B,C\}$ are represented by their instantaneous firing rates $r_X$ which are binned to 1 \unit{ms}. We assume they are well approximated by $r_X \approx \sum_{t=1}^T n_X^{(t)} / 1 \unit{ms}$, where $T=1\unit{ms}/\Delta t$ and $n_X^{(t)},t=1,\ldots,T$ are independent realizations of spike counts (avalanches) $n_X$ observed in $X$. Thus, we get for the correlations between neural activities
\begin{equation}\label{eq:crr}
    C(r_X, r_Y) =
        \frac{\operatorname{cov}(\sum_{t=1}^T n_X^{(t)}, \sum_{t=1}^T n_Y^{(t)})}{\sqrt{\operatorname{var}(\sum_{t=1}^T n_X^{(t)})
        \operatorname{var}(\sum_{t=1}^T n_Y^{(t)})}}
    = \frac{T \operatorname{cov}(n_X, n_Y)}{T\sqrt{\operatorname{var}(n_X)
        \operatorname{var}(n_Y)}} = C(n_X, n_Y)\, .
\end{equation} 
Correspondingly, we obtain for correlations between flicker signal $f_Z$, $Z \in \{A, B\}$ and neural activities $r_X$
\begin{equation}\label{eq:cfr}
    C(f_Z,r_X) = 
        \frac{\operatorname{cov}(f_Z, \sum_{t=1}^T n_X^{(t)})}{\sqrt{\operatorname{var}(f_Z)
        \operatorname{var}(\sum_{t=1}^T n_X^{(t)})}} 
    = \frac{\operatorname{cov}(f_Z, n_X)}{\sqrt{\operatorname{var}(f_Z)\operatorname{var}(n_X)/T}} = \frac{1}{\sqrt{T}} C(f_Y,n_X)\, .
\end{equation}


For evaluating variance and covariance in these correlations, we have to construct the full joint probability distribution $P^\theta(n_A, n_B, m_a, n_C, f_A, f_B)$ from the previously computed distributions $P^\theta_B,P^\theta_A,P^\theta_a$ (Eqs.~\ref{eq:Pthetas}-\ref{eq:Pthetaa}) which are conditioned on avalanches starting either in $A, B$, or $a$. For this purpose we need to consider the event that external input to a selected unit does not lead to an avalanche. This is achieved using~\citep[Proposition F.2]{schunemann2022rigorous} by changing the normalization of the distributions for non-empty avalanches to the phase space volume of the whole system given by $V \coloneqq det( \mathds{1}-W )$, 
where $W$ is the coupling matrix of this model~\eqref{eq:Wfull}\footnote{Note that $V$ can be reduced to an expression involving a $4\times 4$ matrix determinant $\mathcal{V}_4((n, n, m, n)^T, \1, \Wr)$ introduced in~\eqref{eq:vrbs}}.

In the following we will list the non-zero entries of $P(n_A, n_B, m_a, n_C \,|\, f_A, f_B)$.
We have 
\begin{align*}
    P(0, 0, 0, 0 \,|\, f_A, f_B) =
        & 1-p_A-p_B-p_a + \\
        & p_A(1-u_0 P_{A}^{(s)}/V) + p_B(1-u_0 P_{B}^{(s)}/V) + p_a(1-u_0 P_{a}^{(s)}/V)
        \nonumber\\
        =
        & 1-{u_0 \over V}\sum_{X=\{A, B, a\}} p_X P_X^{(s)}
\end{align*}
where the right hand side of the first line is the probability that no unit receives external input, while the terms on the second line reflect the probability that the external input did not push any membrane potential of the receiving units above threshold.
$P^{(s)}_B,P^{(s)}_A,P^{(s)}_a$ are normalization constants needed to apply~\citep[Proposition F.2]{schunemann2022rigorous} 
and defined by equations~(\ref{eq:PV1Bs},~\ref{eq:PV1As},~\ref{eq:PCs}), respectively.

Conversely, we have the following probabilities considering the three possible choices for an avalanche to start in either one of the populations $A,B$ or $a$:
\begin{align*}
    P(n_A, 0, 0, n_C \,|\, f_A, f_B) =
        & (u_0/V) p_A P^{(s)}_A P^{\theta}_A(n_A, n_C) &\text{ if } n_A>0\\
    P(0, n_B, 0, n_C \,|\, f_A, f_B) =
        & (u_0/V) p_B P^{(s)}_B P^{\theta}_B(n_B, n_C) &\text{ if } n_B>0\\
    P(n_A, 0, m_a, n_C \,|\, f_A,f_B) =
        & (u_0/V) p_a P^{(s)}_a P^{\theta}_a(m_a, n_A, n_C) &\text{ if } m_a>0 \, . 
\end{align*}
Taken together, these expressions specify the \emph{joint distribution of flicker values and population activity in all populations} of the model, 
\begin{align}\label{eq:jointdist}
    P(n_A, n_B, m_a, n_C, f_A, f_B) = P(n_A, n_B, m_a,n_C \,|\, f_A, f_B) P(f_A, f_B) \, .
\end{align}
Since the flicker levels are both distributed uniformly and independently on $n_l$ uniformly spaced levels between $-1$ and $1$, each pair of levels is assumed with probability $P(f_A, f_B) = 1/n_l^2$.

\subsection{Correlation analysis of signal routing}

After developing the tools to analytically describe avalanche dynamics and signal correlations in routing network, we can now investigate selective processing in detail.

In order to be successfully routed to V4, the flicker signals $f_A(t)$ and $f_B(t)$ have first to be well represented in the population rates of $A$ and $B$, and then to be transmitted to $C$ and expressed in its population rate which is the output sent to higher visual areas. For investigating signal routing, we first focused on these two steps in isolation, in order to understand later how they act in combination.

For analysis, we denote the instantaneous population firing rates in the networks $a, A, b, B,$ and $C$ binned to $1\unit{ms}$ intervals by $r_a(t), r_A(t), r_b(t), r_B(t),$ and $r_C(t)$, respectively. In order to simplify notation, we usually drop the time index $t$ for the instantaneous firing rates and flicker signals. We quantify signal routing by computing the Pearson temporal correlation coefficients $C(x, y)$ between two time-varying signals $x$ and $y$ as explained in the last subsection, for each of the following each network stages: 
\begin{enumerate}
    \item \emph{Stimulus signal representation} in V1 is measured by the correlations $C(f_A, r_A)$ and $C(f_B, r_B)$ between the flicker signals and the population firing rates in V1.
    \item \emph{Signal transmission} from V1 to V4 is measured by $C(r_A, r_C)$ and $C(r_B, r_C)$.
    \item \emph{Stimulus signal routing} from the flicker signals to V4 is measured by $C(f_A, r_C)$ and $C(f_B, r_C)$.
\end{enumerate} 
In addition to the Pearson correlation, we also report the values of the frequency-resolved spectral coherence $SC(x, y, f)$ between two signals $x$ and $y$ at frequency $f$. Spectral coherence is calculated based on the complex coefficients of the wavelet transforms of the input signals (details see Methods, equation~\ref{eq:defSC}). We calculated the spectral coherence values between pairs of the input signal and the local field potentials $l_A$, $l_B$, and $l_C$ which we approximate in our model by filtering the corresponding population rates with an exponential kernel with time constant $\tau=\unit[15]{ms}$. The reason to use two different correlation measures is feasibility of analytical treatment in case of the Pearson correlation, and comparison with previously published data \citep{grothe2018attention,harnack2015model} as well as having the opportunity to assess the spectral characteristics of routing in case of the spectral coherence.

\begin{figure}[hbt]
    \centering
    \includegraphics[width=\textwidth]{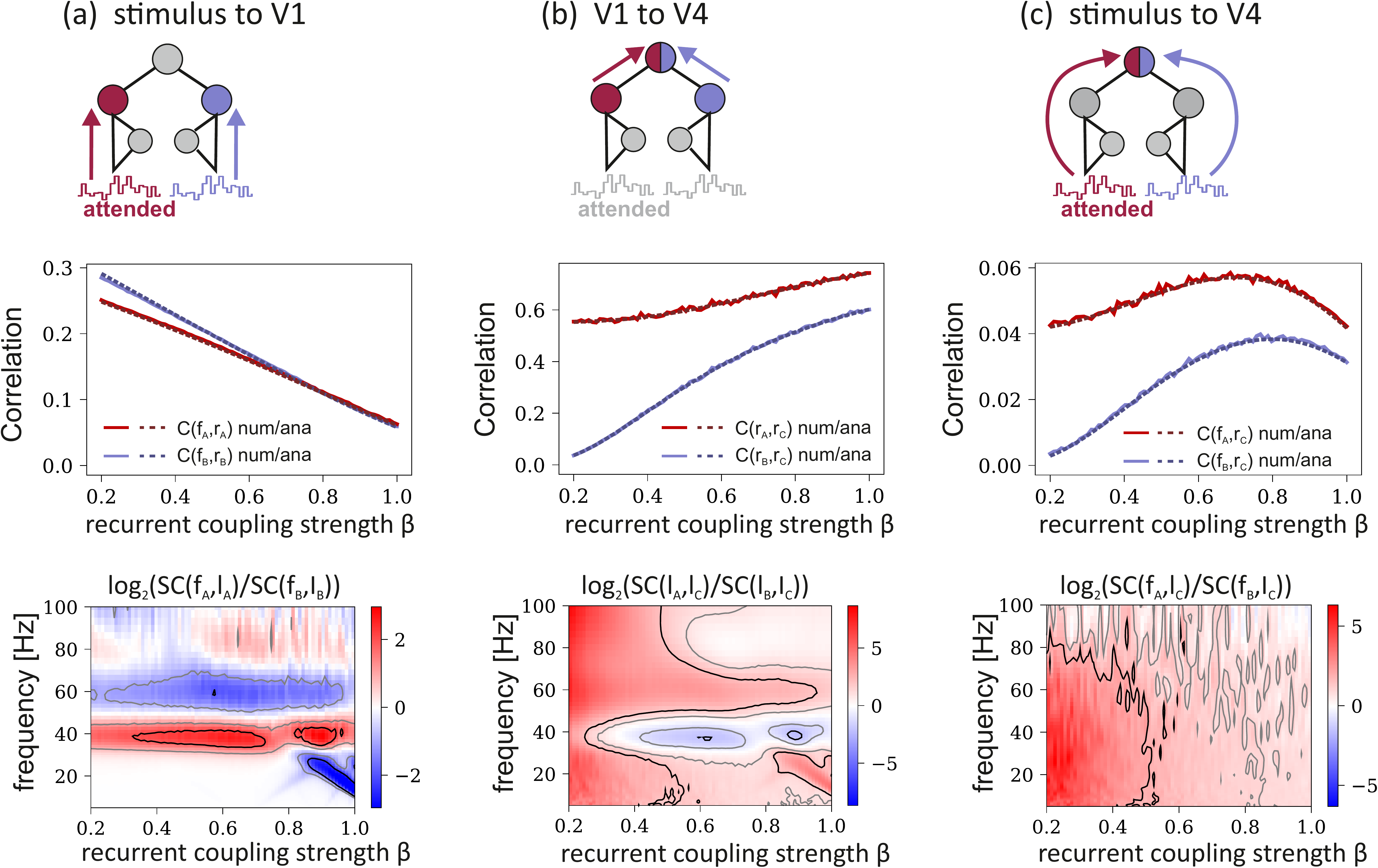}
    \caption{Correlation analysis of signal routing in dependence of recurrent coupling strength.
    Top row: Pearson correlation coefficients in dependence of V1 recurrent coupling strength $\beta$. Dashed lines show analytical results, and solid lines simulation results. Low values of $\beta$ correspond to asynchronous dynamics, while V1 exhibits power law avalanche statistics at $\beta=1$. Blue and red lines mark the results for the nonattended and attended pathways, respectively. 
    (a) Stimulus representation in V1. Correlation between flicker stimuli and corresponding population response decrease with growing $\beta$. The attended signal is slightly less well represented in V1 compared to the non-attended signal. (b) This situation reverses for the correlations between population activity in  V1 and V4. Correlation coefficients grow with $\beta$ and the correlation is higher for the pair $A$ and $C$ compared to $B$ and $C$. (c) For the complete pathway from stimulus to V4, there is a trade-off between the stimulus representation in V1 and subsequent transmission to V4. Optimal correlation values between stimulus and V4 response are assumed at intermediate values of $\beta$ with a large advantage for the attended stimulus. \\
    Bottom row: Ratios between the frequency-resolved correlations (spectral coherences) for the attended and non-attended pathways in dependence of coupling strength. Colors indicates the binary logarithm of the corresponding SC ratios.    Gray and black contour lines mark the regions in which one of the spectral coherence values is at least twice or four times as big as the other one, respectively. }
    \label{fig:correlation-analysis}
\end{figure}

The intra-population coupling $\beta=w_{A,A}=w_{B,B}$ is a control parameter which we can adjust to target a specific dynamical regime or to fit physiological data. Figure~\ref{fig:correlation-analysis} shows the results of the correlation analysis in dependence of the dynamical state of the V1 populations determined by $\beta$. A value of $\beta=1$ corresponds to a critical dynamics, in which population activity is composed of power-law distributed avalanches. Lowering $\beta$ pushes the populations to a more asynchronous irregular dynamics, while $\beta=0$ corresponds to uncoupled neurons, cf. figure~\ref{fig:ehe-rasters-avalanches}. 

For studying the routing network in a physiologially plausible regime, the mean population firing rate of the non-attended population $B$ was kept constant at $\bar{r}_B=\unit[40]{Hz}$ for all $\beta$ values by scaling the external input strength. Since attention is directed to $A$, the control population $a$ is disinhibited and the mean firing rate in $A$ surpasses the rate of $B$ by $30\%$ at $\bar{r}_A=\unit[52]{Hz}$ for the choosen parameters (discussed later).

\textbf{Stimulus signal representation in V1.} First we focus on the stimulus signal representation in V1 which is shown in column (a) of figure~\ref{fig:correlation-analysis}. Both $C(f_A, r_A)$ and $C(f_B, r_B)$ are strictly monotonically decreasing with the recurrent coupling strength $\beta$. This effect arises from the increase of population synchrony in V1 with $\beta$, which leads to a higher variance around the average population firing rate compared to a weakly coupled (or uncoupled) network of neurons. The higher the population synchrony, the more recurrent feedback dominates over feedforward input. In addition, the increase in population synchrony in $A$ induced by the control population $a$ leads to a weaker stimulus signal representation in $A$ compared to $B$, which is most prominent for small recurrent coupling $\beta$.

For assessing the frequency-resolved differences in the stimulus signal representations in populations $A$ and $B$, we show the log-ratio of the corresponding spectral coherences~\eqref{eq:defSC}. The reduction in stimulus signal representation which is suggested by the Pearson correlation is most prominent at high coupling strengths $\beta>0.8$ and in the frequency ranges $f<\unit[20]{Hz}$ and $\unit[45]{Hz}<f<\unit[65]{Hz}$, but there is also enhancement in the $\unit[40]{Hz}$-range.

\textbf{Signal transmission from V1 to V4.} The column (b) of figure~\ref{fig:correlation-analysis} shows signal transmission from $V1$ to $V4$. In contrast to (a), the correlation coefficients $C(r_A, r_C)$ and $C(r_B, r_C)$ grow monotonously with $\beta$ and optimal signal transmission is attained if the V1 populations operate in their critical states. The added synchrony in population $A$ leads to higher correlations $C(r_A, r_C) > C(r_B, r_C)$. For sufficiently small values of $\beta$, the V1 population $B$ processing the non-attended signal loses the ability to drive V4 population $C$ almost completely, which leads to $C(r_B, r_C)\approx 0$. 

Correspondingly, for all frequencies and for low values of $\beta$, the ratio of spectral coherences $SC_{l_A, l_C}(f)/SC_{l_B, l_C}(f)$ between attended V1 -- V4 and non-attended V1 -- V4 signal transmission are above $1$. For higher values of $\beta$ and in the range $\unit[35]{Hz}<f<\unit[45]{Hz}$, the ratio of the spectral coherences decreases below $1$, while it still stays above $1$ for all other frequencies. The advantage for the population $A$ processing the attended stimulus to shape the output of $C$ 
is highest in the frequency range $\unit[45]{Hz}<f<\unit[65]{Hz}$ by a factor bigger than four.

\textbf{Stimulus signal routing from flicker to V4.} Combining these two steps yields routing from the stimulus flicker signal to the output activity in V4 population $C$, which is shown in the third column of figure~\ref{fig:correlation-analysis}. Maximal values of selective routing $C(f_A, r_C)$ and $C(f_B, r_C)$ are attained for \emph{intermediate} values of $\beta$, due to the required balance between the objectives of having a good signal representation in V1 and a good transfer to V4. While the ratio of firing rates between the attended V1 populations $A$ and $B$ is pushed to $1.3$ by the disinhibited control population $a$, the increase in population synchrony in $A$ leads to ratios far above $1.3$ for the correlation coefficients measuring signal routing of $f_A$ to V4 when compared to $f_B$. This supralinear effect is present even stronger in the ratios of spectral coherences, where for all frequencies we find $SC_{f_A, l_C}(f) > SC_{f_B, l_C}$. In general, across frequencies the ratios decrease with $\beta$. At $\beta=0.8$ they are at a mean value of around two which fits to physiologically observed values \citep{grothe2018attention}.


\subsection{Reproducing physiological signatures of selective signal processing}

After describing the dependence of correlation based routing measures on the control parameter $\beta$, let us zoom in to the parameter value $\beta=0.76$, for which the correlation $C(f_A, r_C)$ is maximized (c.f. figure~\ref{fig:correlation-analysis} panel c, red curve) and compare signatures of selective information routing in the model to signatures of selective visual attention identified in previous experiments. Afterwards, we will show that the qualitative nature of the observed attentional modulations generalize to a larger range of parameter values in the routing network.

\begin{figure}[hbtp]
\centering
\includegraphics[width=\textwidth]{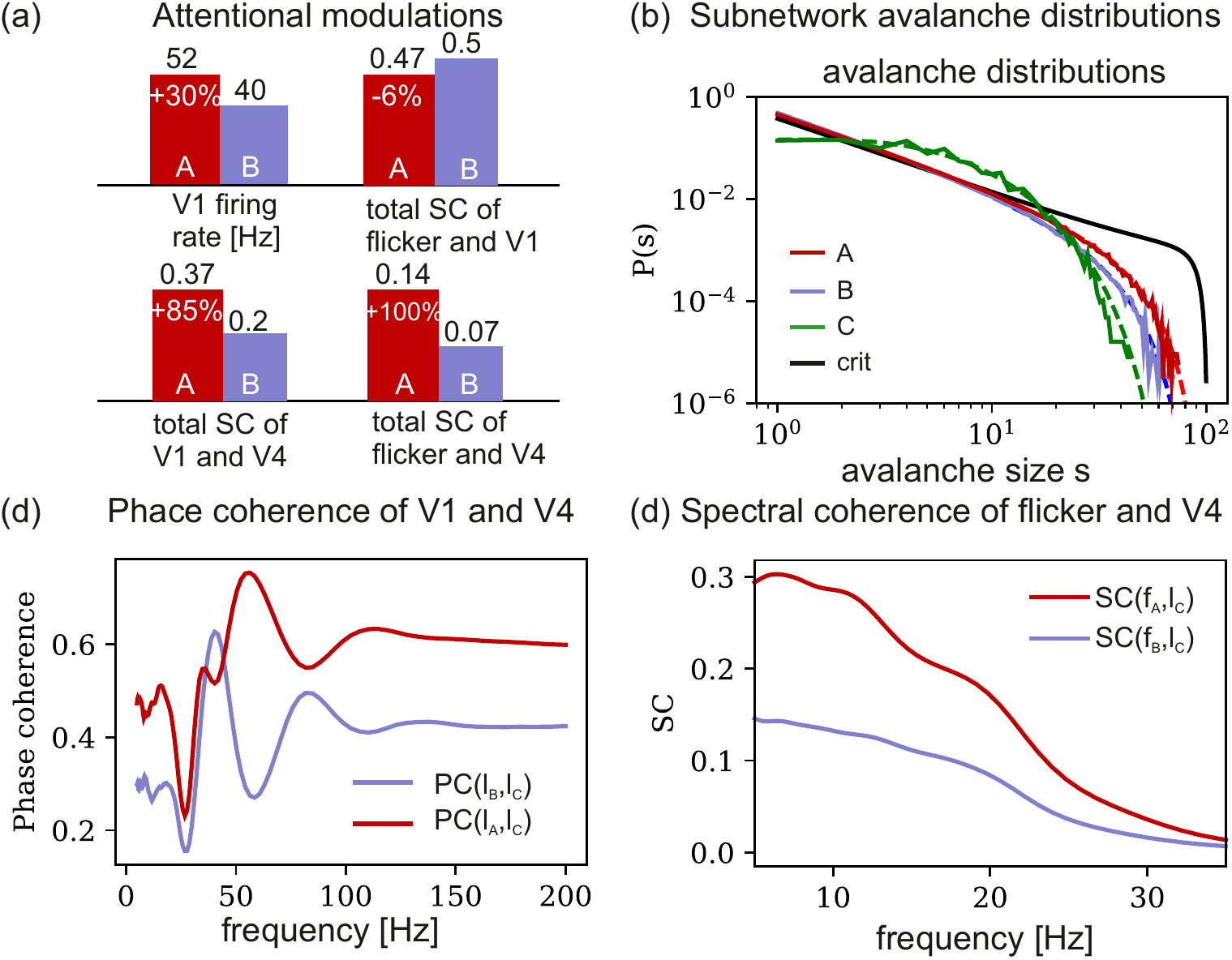}
\caption{Detailed analysis of signal routing for the model parameters  $\beta=0.78,\theta=5$.\\
(a) Summary of effects on firing rate and mean spectral coherences for the frequency range shown in figure~\ref{fig:correlation-analysis} for all stages of the model.
(b) Numerical and analytical avalanche distributions of populations $A$, $B$, and $C$. Disinhibiting the control population leads to a shift towards criticality in the distribution for V1 population $A$.
(c) Phase coherence between V1 and V4. At the largest peak of the red curve, the phase coherence between the V1 population processing the attended signal and V4 is higher by a factor of 2.76 than the corresponding phase coherence for the non-attended case.(d) Spectral coherence between stimulus and V4 shows  preferential routing of $f_A$ for all frequencies with an advantage over $f_B$ by a factor of around two.}
\label{fig:prop-opt}
\end{figure}

Figure~\ref{fig:prop-opt} summarizes the effect sizes of attentional modulation, the subnetwork avalanche distributions and a frequency-resolved analysis of signal routing for the value of $\beta=0.76$. The firing rate advantage for the V1 population $A$ processing the attended stimulus over V1 population $B$ is $30\%$ (panel A). This effect size is well within the observed effect sizes of attentional rate modulations in area V1 for two competing stimuli of matching contrast reported in~\citet{SchuenemannM2021e}. The correlations between the attended flicker modulations and the V4 population $C$ increase at the same time by $60\%$. However, the values of spectral coherences between the flicker signals and the corresponding V1 populations, summed over frequencies, \emph{decrease} to $\sum_fSC_{f_A, l_A}(f) \approx 0.9 \sum_f SC_{f_B, l_B}(f)$. This is consistent with the result of a control experiment reported in~\citep{grothe2018attention}, and in our model is due to the weakened signal representation by the increased population synchrony of V1 population $A$ as described previously.


At the same time the increase in population synchrony in V1 population $A$ leads to a shift in its avalanche distribution towards criticality. Panel (b) of figure~\ref{fig:prop-opt} shows the distributions of sizes of avalanches in the populations of the model generated from $3750$ seconds of numerical simulation together with the analytical avalanche size distributions given by Eq.~\ref{eq:avsV1B},\ref{eq:avsV1A},\ref{eq:avsV4}. Numerical and analytical results are in perfect agreement. Compared to the avalanche size distribution of V1 population $B$, the tail of the distribution in $A$ increases significantly. However, comparison with the avalanche size distribution at criticality (black line, $\beta=1$) shows that population $B$ is still fairly subcritical.

It was shown that not only the synchrony of the attended V1 populations (measured by power in the gamma band) increases by attention, but also the (phase) synchronization between $V1$ and $V4$ -- sometimes even by a factor as large as four~\citep{grothe2012switching}. 
Panel (c) of \Fig{fig:prop-opt} shows the phase coherences $PC_{l_A, l_C}$ and $PC_{l_B, l_C}$ computed via \Eq{eq:defPC}. The largest phase synchronization is attained between V1 population $A$ and V4 population $C$ in the frequency band between $\unit[50]{Hz}$ and $\unit[70]{Hz}$, and it is larger by a factor of $2.76$ at the peak frequency. 
In a narrow frequency range below the beak frequency there is a region where the phase coherence to the V4 local field potential is slightly higher for the non attended V1 population $B$ population. Such a slight attenuation of phase coherence has also qualitatively been observed in experimental studies \citep{bosman2012attentional}.
 
Finally, tagging the stimuli with independent realizations of a broadband flicker modulation allowed Grothe et al.~\citep{grothe2018attention} to compute the spectral coherence of the shown luminance modulations with the measured V4 LFP. They reported that spectral coherence between the flicker signal added to the attended stimulus and the LFP recorded in V4 was elevated in comparison to the spectral coherence with the flicker signal added to the non-attended stimulus. This effect occurred for all frequencies up to 20~--~25\unit{Hz} and amounted to an average two-fold increase in spectral coherence when summed over frequencies, $\sum_f SC_{f_A,l_C}(f) \approx 2 \sum_f SC_{f_B,l_C}(f)$. This is consistent with the relation of spectral coherences in our model, as shown by panel (d).

\subsection{Effect of synchronicity gain}

In the previous subsections we have demonstrated that our model is capable to reproduce a wide range of physiological signatures of selective signal routing across different observation levels in one coherent framework and within one single parameter set. In this section we will discuss how the signal routing mechanism in the model depends on the network 
parameters, especially the parameter $\theta$ which regulates the strength of the synchronicity gain in V4. The synchronicity gain models the relative increase in V4 response to synchronous input, compared to input of the same total strength but randomly distributed over a larger time interval.

In our model, $\theta$ acts as a \emph{threshold} for the connections from V1 populations $A$ and $B$ to V4 population $C$. The V4 population $C$ is only driven if at least $\theta$ neurons are jointly active in one of the V1 populations, i.e. they elicit an avalanche of at least $\theta$ neurons. According to the probability distribution $P$ of joint population activity given by \Eq{eq:jointdist}, the mean firing rates of a neuron in population $Y$ is equal to
\begin{equation}
    \bar{r}_Y = {1 \over n\Delta t} \sum_{s=1}^n s \, P(n_Y = s) \, .
\end{equation}
With $\theta>1$, the drive of the V4 population is not proportional to the population rates $\bar{r}_A+\bar{r}_B$ anymore, but to the \emph{tail} rates given by 
\begin{equation}
    \bar{r}_Y^\theta = {1 \over n\Delta t} \sum_{s=\theta}^n s \, P(n_Y = s) \, .
\end{equation}
The tail rates determine the ratio of correlation coefficients between the flicker signals and V4 activity in our model,
\begin{align}
    \label{eq:routing-selectivity}
    \frac{C(f_A, r_C)}{C(f_B, r_C)} = \frac{\bar{r}_A^\theta}{\bar{r}_B^\theta} \, .
\end{align}

A consequence of this equation is that for $\theta=1$, the routing selectivity would just be the ratio of average firing rates $\bar{r}_A/\bar{r}_B$ and thus there would be no supralinear effect on the signal routing.

The same limiting case would be reached if the extra external input by the disinhibited control population would increase the firing rate of V1 population $A$ without increasing its population synchrony i.e. if the neurons in $a$ would not be connected recurrently.

\begin{figure}[hbtp]
\centering
\includegraphics[width=\textwidth]{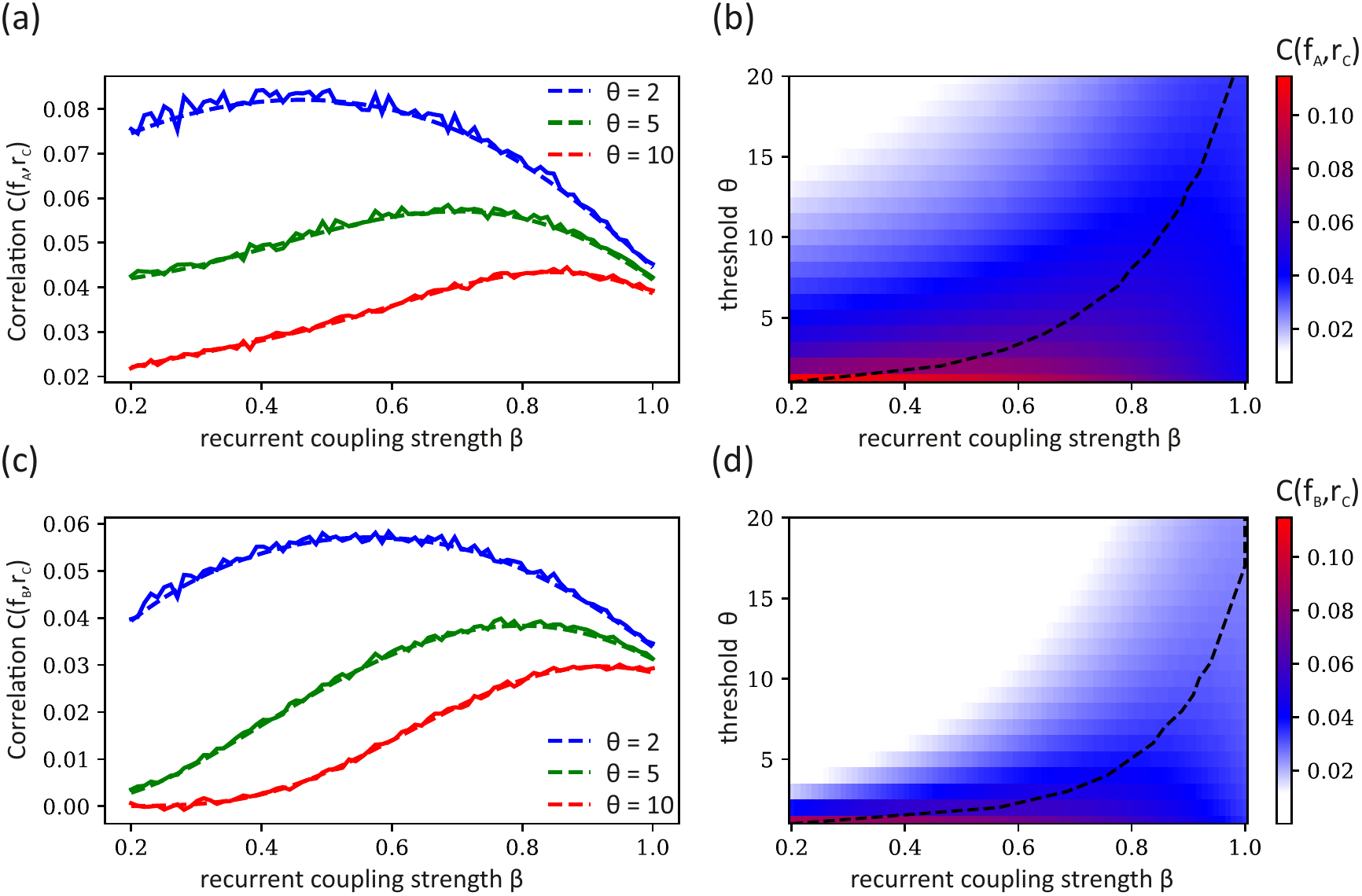}
\caption{Effect of synchronicity gain. (a) and (c): Numerical (solid lines) and analytical (dashed lines) correlation coefficients $C(f_A,r_C)$ measuring signal routing of the attended stimulus to V4 (a) and $C(f_B,r_C)$ measuring signal routing of the non-attended stimulus to V4 (c) in dependence of coupling strength $\beta$ for synchronicity threshold values $\theta=2$ (blue), $\theta=5$ (green), and $\theta=10$ (red). Correlation coefficients decrease with growing $\theta$ and have a unimodal shape for fixed $\theta$ in dependence of $\beta$, with the location of the peak moving to higher values of $\beta$ with growing $\theta$. Panels (c) and (d) show the analytical correlation coefficients for all synchronicity threshold values up to 20 for the attended and non-attended signal, respectively. For each $\theta$, the dashed lines mark the coupling $\beta$ for which correlations become maximized.}
\label{fig:correlation-phase-space}
\end{figure}

How does the signal routing performance of our model depend on the strength of the synchronicity gain of $V4$ for values $\theta>1$? The left column of \Fig{fig:correlation-phase-space} shows the correlations $C(f_A, r_C)$ (top) and $C(f_B, r_C)$ (bottom) in dependence of $\beta$ for three values $\theta=2,5,10$.  Analytical correlation coefficients are shown in dashed lines and are again in perfect agreement with the numerical results. Since the $\theta$ 'tail rates' which drive the V4 population $C$ become smaller, the correlation coefficients decrease with growing $\theta$. In addition, the location of maximal correlation values shift to larger $\beta$ values for higher thresholds $\theta$. These effects hold for all values of $\theta$ as is shown in the analytical correlation values $C(f_A, r_C)$ (top) and $C(f_B, r_C)$ (bottom) shown in the right column of \Fig{fig:correlation-phase-space} in dependence on both $\beta$ and $\theta$. The dashed black line marks the value $\beta$ of the control parameter maximizing the correlation between flicker and V4 activity for each value of the synchronicity threshold $1\leq \theta \leq 20$. 

These results indicate a trade-off between routing \emph{performance} measured by the correlation between the flicker modulation and the V4 population activity, and the routing \emph{selectivity} measured by $C(f_A, r_C)/C(f_B, r_C)$. While the selectivity grows (without bounds) for $\beta\rightarrow 0\,,\theta \rightarrow n$, the routing performance has for each value of $\theta$ a global maximum. The connection values $\beta$ for which these maxima are observed are connected by the dashed lines in the right column of \Fig{fig:correlation-phase-space}. Note that the maximally attainable routing performance decreases with $\theta$.

\subsection{Signal routing to leaky integrate and fire neurons}

Since the implementation of the synchronicity gain in our model is somewhat artificial, we studied the generality of our findings by driving a (dimensionless) leaky integrate and fire neuron which follows the differential equation $\tau dV/dt = -V + \Delta V\sum_{i}s_i\delta(t_i)$  with the activity of V1 populations $A$ and $B$, where $\tau=\unit[3]{ms}, \Delta V=0.02$ and where incoming avalanches of sizes $s_i$ at times $t_i$ lead to an instantaneous increase in $V$ by $s_i \Delta V$. The firing threshold is set to one, and if a neuron crosses that threshold its membrane potential is immediately reset to $V=0.$ 

With this choice of parameters, the integrate and fire neuron is tuned to operate in a coincidence detection mode, such that it responds with a higher firing rate according to the level of synchrony in the incoming spikes when driven by the population activity of V1 populations $A$ and $B$. We denote the firing rate of the integrate and fire neuron binned to $\unit[1]{ms}$ intervals by $r_{\text{iaf}}$. 

\begin{figure}[hbtp]
    \centering
    \includegraphics[width=0.9\textwidth]{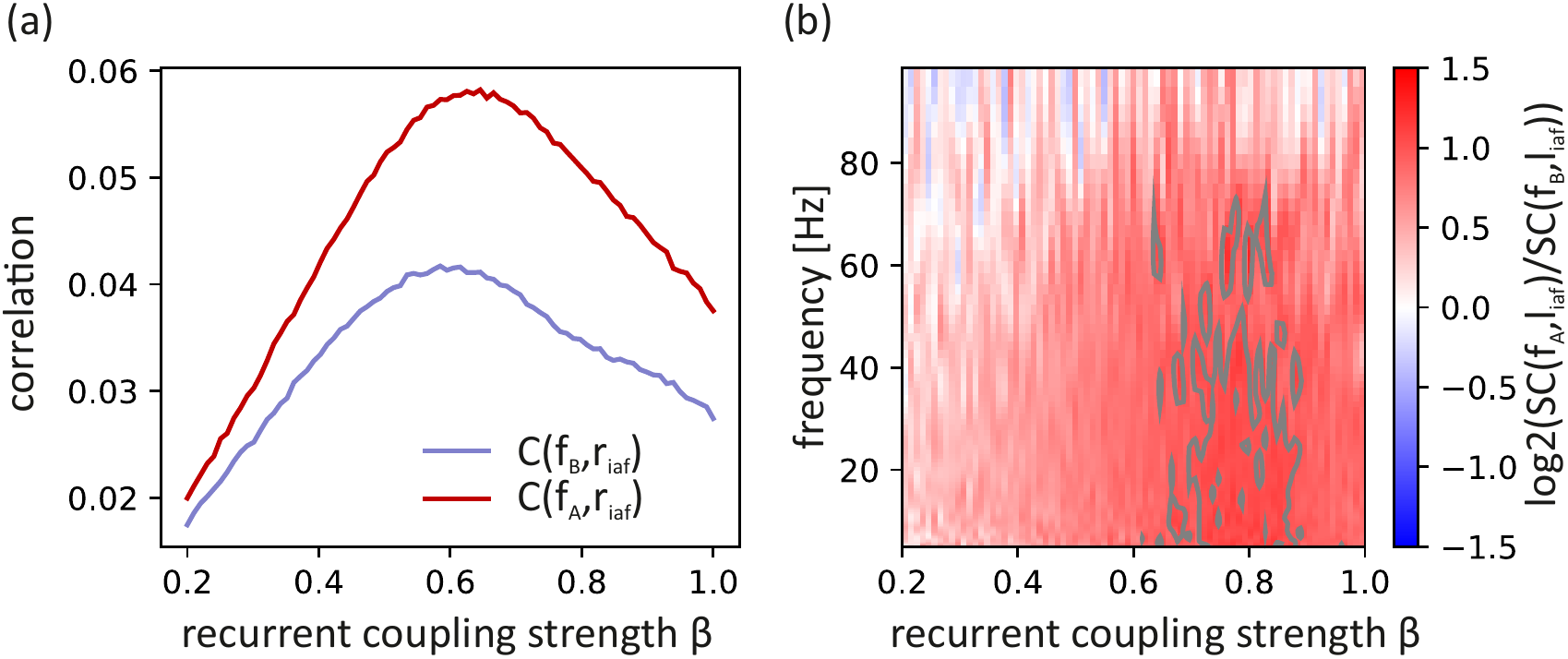}
    \caption{Signal routing to a single leaky integrate and fire neuron. 
    Parameters of the integrate and fire neuron are $\tau=\unit[3]{ms},V_{\mathrm{thr}}=1,V_{\mathrm{rest}}=0$, and $\Delta V=0.02$ where $\Delta V$ is the instantaneous increase in membrane potential caused by a single incoming spike. (a) Pearson correlation coefficients between flicker signals and firing rate of the leaky integrate and fire neuron. (b) Binary logarithm of the ratio of spectral coherences between the attended and non-attended signal and the instantaneous firing rate of the integrate and fire neuron, filtered with a $\tau=15\unit{ms}$. Gray contour lines mark regions in which the spectral coherence ratio is at least two.}
    \label{fig:iaf-res}
\end{figure}

\Fig{fig:iaf-res}, panel (a), shows that the resulting correlation coefficients in dependence of $\beta$ show a qualitatively similar, uni-modal shape like the flicker to V4 correlations in \Fig{fig:correlation-analysis}. For low values of $\beta$, both the attended and non-attended V1 populations are basically unable to drive the V4 population, leading to low values of both correlation coefficients. This differs from the purely EHE-neuron based model analyzed previously. I Figure~\ref{fig:correlation-analysis} (c), the correlation of the attended signal and V4 activity (red) curve does not decay to zero for $\beta \rightarrow 0$. This is a result of the chosen hard synchronicity threshold $\theta=5$, which still allows the synchronicity inherent in the activity of the control population on the attended pathway to drive V4 even if neurons in the corresponding V1 population would be uncoupled.

The selectivity of signal routing to the integrate and fire neuron is optimized around the location of the peak in correlation $C(f_A,r_{\text{iaf}})$ at a recurrent coupling strength of around $\beta=0.8$, being more than twice as strong as the corresponding coherence to the non attended signal.

\subsection{Information theoretic analysis of signal routing}

While the correlation (and spectral coherence) values are easy to obtain from simulations and were used in experimental investigations like~\citep{grothe2018attention}, they are only designed to measure linear dependencies. The strong nonlinearities of brain dynamics thus call for more general information theoretic measures.
We argued 
that the \emph{unique information} between the flicker signals and V4 activity is particularly well suited for this problem. Unique information is one realization of Partial Information Decomposition (PID) which extends classical information theory to decompose the mutual information between an 'output' variable and two 'input' variables into non-negative values of shared information, synergistic information and two unique information values. 

In this section, we denote the mutual information between two discrete random variables $X,Y$ by $I(X,Y)$ defined as
\begin{align*}
I(X,Y) = \sum_{x} \sum_{y} p_{(X,Y)}(x,y) \log \left ( \frac{p_{(X,Y)}(x,y)}{p_X(x)p_Y(y)}\right ) \text{ ,}
\end{align*}
following the convention $0\log(0) = 0$.

Applied to our model, PID decomposes the mutual information $I((f_A,f_B),n_C)$  between the pair of flicker levels $(f_A, f_B)$ and the V4 activity $n_C$ into the four components
\begin{enumerate}
\item \emph{unique information} $UI(n_C : f_A\setminus f_B)$ of $f_A$ about $n_C$ which is not included in $f_B$,
\item \emph{unique information} $UI(n_C:f_B \setminus f_A)$ of $f_B$ about $n_C$ not included in $f_A$,
\item \emph{shared information} $SI(n_C:f_A;f_B)$ about $n_C$ which is redundantly encoded in both $f_A$ and $f_B$,
\item and the \emph{synergistic information} $CI(n_C:f_A;f_B)$ (also called common information) which is only accessible from the joint distribution of $f_A,f_B,n_C$.
\end{enumerate}

There are three equations which relate the four terms of the Partial Information Decomposition to mutual information values
\begin{align*}
    I(n_C, f_A) = & UI(n_C: f_A \setminus f_B) + SI(n_C: f_A; f_B) \\
    I(n_C, f_B) = & UI(n_C: f_B \setminus f_A) + SI(n_C: f_A; f_B) \\
    I(n_C, (f_A, f_B)) = & UI(n_C: f_A \setminus f_B) + UI(n_C: f_B \setminus f_A) + \\
        & SI(n_C: f_A; f_B)+CI(n_C: f_A; f_B)
\end{align*}
These three equations do not uniquely specify the four non-negative values of the PID and since the introduction of PID by~\citep{WilliamsBeer:Nonneg_Decomposition_of_Multiinformation}, multiple different measures were introduced. Here, we focus on the measure introduced by Bertschinger et. al~\citep{bertschinger2014quantifying} which has desirable mathematical properties but is hard to evaluate in practice which involves solving a convex optimization problem. 

\begin{figure}
    \centering
    \includegraphics[width=\textwidth]{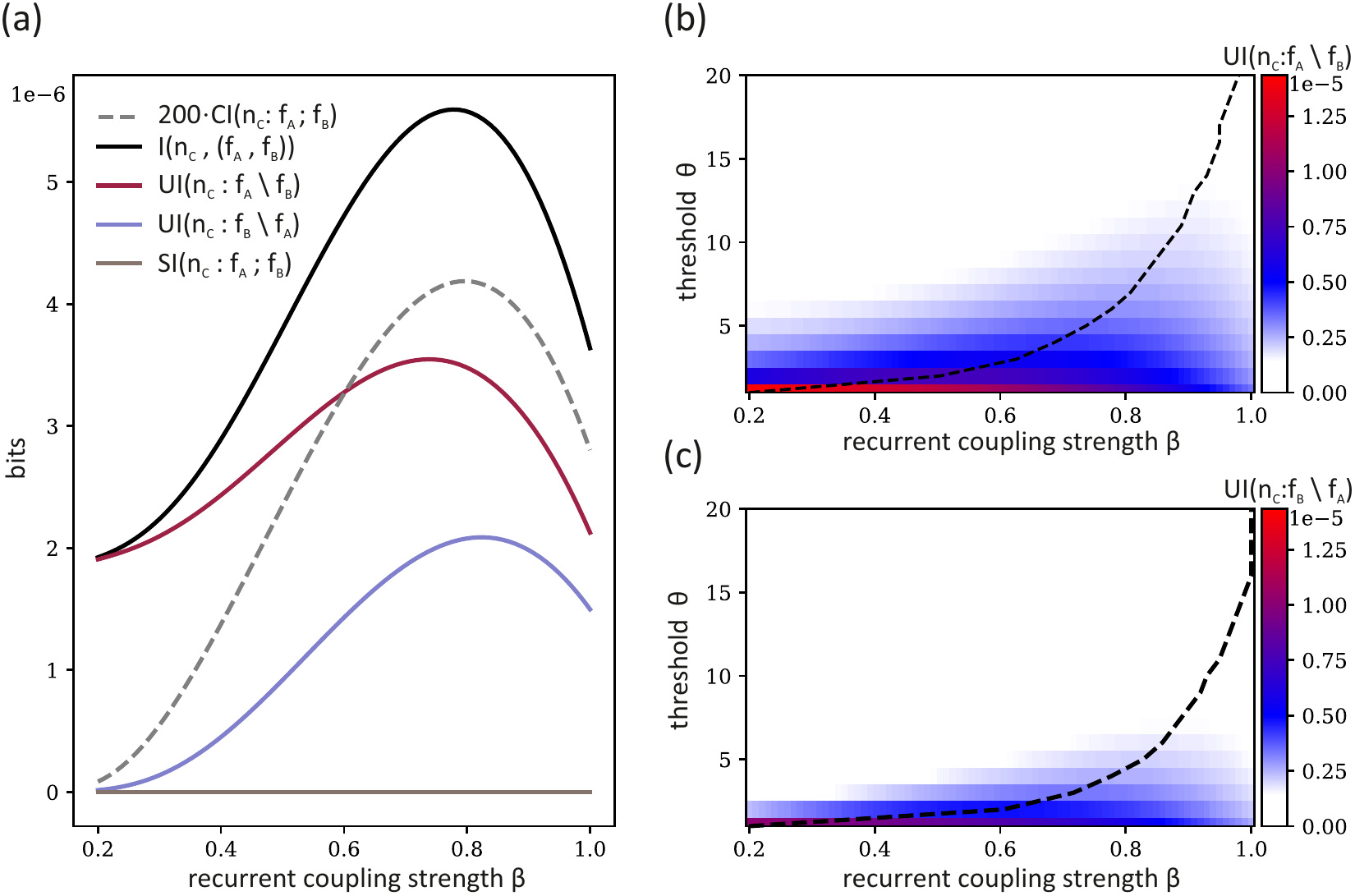}
    \caption{Partial information decomposition of signal routing. (a) Mutual information $I(n_C,(f_A,f_B))$ between the pair of inputs signals and the number of spiking units in V4 population $C$ (black curve) together with its partial information decomposition into unique information ($UI(n_C:f_A\setminus f_B)$: red curve, $UI(n_C:f_B\setminus f_A)$: blue curve), shared information ($SI(n_C:f_A;f_B)$: brown curve) and synergistic information ($CI(n_C:f_A;f_B)$: gray dashed curve, scaled by a factor of 200) in dependence of $\beta$ for $\theta=5$. The decomposition shows a predominant signal routing mode characterized by high unique information values, completely vanishing shared information, and synergistic information which is two orders of magnitude lower than the unique information values. Similar to the corresponding correlation values, the unique information values have a unimodal shape with maxima in the vicinity of the maxima found for the correlation results shown in Figure~\ref{fig:correlation-analysis}, panel c. (b) and (c): Heat plots of unique information values about $n_C$ of the attended signal $UI(n_C:f_A\setminus f_B)$ (b) and the non-attended signal $UI(n_C:f_B\setminus f_A)$ (c) in dependence of the V1 recurrent coupling strength $\beta$ and synchronicity threshold $\theta=1 \ldots 20$. Dashed lines mark the maxima of unique information in each row and show the same characteristic shift to higher $\beta$ values for growing $\theta$ as the correlation values in figure~\ref{fig:correlation-phase-space}.}
    \label{fig:pid}
\end{figure}

Note that since $f_A$ and $f_B$ are independent by construction, one would suspect that they possess no shared information about $n_C$. While this criterium alone does not imply vanishing of shared information, 
it turns out that for the joint distribution of $f_A, f_B, n_C$ in our model according to \Eq{eq:jointdist} fulfills the conditions of ~\citep[Lemma 9]{bertschinger2014quantifying}\footnote{This was checked numerically for all shown parameter values.} and thus there is no shared information, $SI(n_C: f_A, f_B) = 0$. With this additional condition, the remaining three terms of the PID follow from the three equations above to be given by classical information theoretic terms which can be simply calculated analytically from the distribution \Eq{eq:jointdist}:
\begin{align*}
    SI(n_C: f_A; f_B) &= 0\\
    UI(n_C: f_A \setminus f_B) &= I(n_C, f_A)\\
    UI(n_C: f_B \setminus f_A) &= I(n_C, f_B)\\
    CI(n_C: f_A; f_B) &= I(n_C, (f_A, f_B))-I(n_C, f_A)-I(n_C, f_B)
\end{align*}
The partial information decomposition of the signal routing for $\theta=5$ is shown in panel (a) of \Fig{fig:pid}. As expected, we find that our model is in a predominantly information routing mode which is characterized by large values of the unique information, with the unique information for the attended signal being consistently higher for all recurrent coupling strengths $\beta$. There is only a small part of synergistic information present, which is smaller by a factor of roughly 200 compared to the unique information values.

The analysis of the signal routing using the unique information measures reproduces the results using Pearson's correlation coefficient qualitatively: Absolute values of the unique information measures shrink with growing synchronicity threshold $\theta$. For a fixed $\theta$ they have a unimodal shape in dependence of coupling $\beta$. The $\beta$ value maximizing the corresponding unique information grows with increasing synchronicity threshold $\theta$. The qualitative similarity between the routing performance measured by correlations and by unique information becomes apparent by comparing the right sides of \Fig{fig:correlation-phase-space} and \Fig{fig:pid}. Here we also see the same trade-off between routing performance, i.e. the unique information about the attended signal reaching V4, and the routing selectivity which can be captured by the ratio of the two unique information values. While the selectivity is very strong for small couplings $\beta$, less information is transmitted in total compared to larger values of $\beta$.

\pagebreak
\section{Discussion}

Here we showed that attention-dependent enhancement of intra-population synchrony establishes a preferential communication channel and reproduces key physiological signatures of selective signal processing. In presence of a synchronicity gain mechanism in the receiving population, the combined effect of firing rate increase and increase in the synchrony of population activity lead to a supralinear effect on signal routing performance. For achieving such enhanced synchrony, we postulate a biophysically realistic mechanism in which attention disinhibits a small control population operating in a critical state, which in turn provides modulatory input to the V1 population processing the attended stimulus thus increasing its synchrony. 


The advantage of our approach is the choice of a model framework which allows for a rigorous mathematical analysis of spontaneous synchronization in modular networks \citep{schunemann2022rigorous}. We were thus able to analyze signal routing performance using both correlation and unique information measures which provided similar qualitative results. It turned out that in order to route input signals to V4, the network has to balance two competing tasks: First, the visual input signal has to be represented well in the V1 activity which is best done with an asynchronous irregular population dynamics. Second, this signal has to be routed from V1 to V4 for which a population activity operating at the critical point is optimal. Combining these requirements, optimal signal routing configurations were obtained neither at a fully asynchronous nor at a critical dynamics, but at intermediate coupling strengths. A second compromise was revealed by quantifying the consequences of varying the level of synchronicity gain on routing: Higher levels of synchronicity gain shifted the stimulus-response correlation maxima to larger values of recurrent coupling $\beta$ values, and increased the selectivity of information routing, but decreased the maximal attainable correlation values. Qualitatively similar results were obtained when replacing the V4 network and its rather abstract synchronicity gain mechanism chosen for allowing mathematical analysis by a biophysically more realistic leaky integrate and fire neuron. These neurons also exhibit a synchronicity gain that is maximized when they are on average driven to be close to their firing thresholds.

The analysis reveals that the proposed routing mechanism does not depend on a particular choice of neuron model but on two aspects: Attention increases the level of intra-population synchrony in the population processing the attended stimulus, and the receiving populations respond stronger to synchronous input as compared to input which is spread more evenly over time and contains the same number of presynaptic spikes. 

While the network model \emph{per se} is not designed to generate oscillations, they nevertheless arise due to recurrent feedback with their frequency depending on interaction strengths and average external drive. Higher levels of population synchrony lead both to a shift of the avalanche size distribution towards a power law (cf.~\ref{fig:prop-opt}, panel B), and at the same time to higher oscillatory power. While particular aspects of the power spectrum of this model may differ from power spectra of populations recorded in vivo, the modulatory effects of attention can be reproduced quantitatively. It is interesting to note that the nonlinear routing effects quantified by the Pearson correlation coefficients and PID analysis do not rely on oscillations at all -- the corresponding measures would not change if we destroyed temporal correlations in the model dynamics by sampling at each timestep i.i.d. realization of the joint distribution of population activity~\eqref{eq:jointdist}.
This observation leads to an important insight: routing-by-avalanches does not require oscillatory activity for selective signal routing, neither in the sending nor in the receiving population -- even if the emergence and observation of an oscillatory dynamics is inseparably linked to such a cortical network.



In this model we could apply the theory of spike pattern formation in recurrent networks developed in~\citet{schunemann2022rigorous} for calculating the joint probability distribution of population activity over all subnetworks in \Eq{eq:jointdist}. By marginalizing this distribution, subnetwork avalanche size distributions were obtained and routing measures could be calculated analytically. 
This model thus constitutes a prime showcase of the mathematical framework demonstrating that it can provide analytical insights into structured networks successfully reproducing signatures of relevant computational tasks. In our case, the population activity allowed us to calculate the avalanche size distributions of all populations and also to compute the correlation coefficients on all stages of the model (eqs.(\ref{eq:avsV1B},~\ref{eq:avsV1A},~\ref{eq:avsV4},~\ref{eq:crr})). Comparison with numerical data showed perfect agreement. This allowed us to understand the influence of model parameters - especially the recurrent coupling strength $\beta$ in $V1$ and the synchronicity gain threshold $\theta$ and provided a closed form expression~\eqref{eq:routing-selectivity} for the selectivity of signal routing.

Using the exact joint distribution of population activity we were able to study the partial information decomposition of~\cite{bertschinger2014quantifying} of the signal routing in the model without the inherent difficulties in applying it to time series data. We showed analytically that there is no shared information about the number of spiking units in V4 between the flicker realisations, and that the three remaining parts of the decomposition can be deduced from classical mutual information values. This circumvented the need to numerically solve a challenging high dimensional complex optimization problem in order to calculate the PID according to~\cite{bertschinger2014quantifying}. As expected from the construction of a predominantly signal routing configurations, the mutual information between the input pair of flicker realizations and V4 activity is mostly composed of the two unique information values with a synergistic information content that is two orders of magnitude smaller. The unique information values reproduced all qualitative features of the correlation analysis, including the dependency of the optimal routing configurations on the synchronicity gain $\theta$. The low synergistic values arise since an avalanche in V4 is either caused by an avalanche from population $A$ or from population $B$, by construction of the external drive which targets only one unit in each of the very small time steps (time scale separation).
Thus, the situation analyzed here would more closely resemble the activity at the the input layer of an in-vivo population in V4, whereas outgoing signals could correspond to results of computations within V4 and thus would contain higher amounts of synergistic information.

From a dynamical systems perspective, routing-by-avalanches (RBA) is much simpler than 'communication-through-coherence' (CTC) schemes which require to coordinate phase relationships between oscillating networks \citep{fries2015rhythms}. For example, the model proposed by \cite{harnack2015model} postulates that V1 and V4 populations engage in independent gamma oscillations. By lateral inhibitory couplings, the V1 populations go into an anti-phase synchronization, while each of them seeks to entrain the receiving V4 population. A rate increase on the V1 population driven by the attended stimulus breaks the symmetry in this bi-stable system and establishes a stable favourable phase relationship of the V1 population sending the attended signal with the receiving V4 population. Although this realization of the CTC scheme is particularly elegant since suitable phase relationships emerge by means of a self-organization process triggered by a simple rate increase to the attended pathway, it is nevertheless a complicated system which needs proper coupling parameters and well-adjusted connection delays to function. In contrast, the RBA model stays in an ergodic equilibrium state in which the $V4$ activity becomes a mixture of the synaptic drives provided by the $V1$ populations. Attention modulates driving strengths and thereby also the signal routing correlations. It is also trivially extendable to selecting signals from one of more than just two V1 populations projecting to a V4 population, since it does not have the problem to be able to put all 'non-attended' populations with mutual frustrating interactions into the same, anti-phase relation with the 'attended' population.

For future work, it will be essential to understand in how far RBA can explain selective processing in the 'real' brain. The observation that our routing network can reproduce key findings is already a good start. As a next step, it would be useful to quantify how much stimulus information synchronous events carry with and without attention. When going along with oscillations, it is likely that large events will be concentrated at favorable phases for signal transfer. At these phases, it has already been shown that spectral coherence, a measure for signal routing, is maximized \citep{lisitsyn2020visual}. Another indication for RBA could be that avalanche distributions exhibit heavier tails under attention -- however, we would still expect that the whole system is not necessarily close to a critical state, since optimizing selective processing requires a balance between the critical and asynchronous random state.
Further implication of the existence of control populations is that additional nodes -- possibly embedded in other layers in V1/V2, become active. In particular, these units would be preferential seeds for avalanches comprising also neurons which are normally active without attention.

Further support to these ideas is given by a recent finding that selective attention is able to overcome even larger rate imbalances induced by stimulus configuration (i.e. having a lower contrast target next to a high contrast distracter) \citep{SchuenemannM2021e}. It appears that over a large working range, attention always establishes a rate advantage for the target stimulus in the sending population. However, in many situations that rate advantage is extremely small compared to the quasi all-or-none effect of selective routing to V4. An explanation of this puzzle might be that the small rate advantage is accompanied by a large synchronicity enhancement in the V4 input drive, which would establish stable RBA. It is even possible that enhanced synchronicity might compensate a (small) rate \emph{disadvantage} for the target stimulus.

In summary, we presented a coherent paradigm for selective processing based on spontaneous synchronization and avalanche propagation. The associated mathematical framework is sufficiently flexible to allow extension from information routing to information processing, for example by embedding the statistics of visual feature conjunctions into the currently homogeneous coupling matrices. For assessing these functions quantitatively, we adapted various statistical methods such as correlation measures and unique information to our framework, which provide a toolset which is also applicable to physiological data.

\clearpage

\section{Methods}\label{sec:model-methods}

\subsection{Network model}

We use the well-studied EHE model~\citep{eurich2002finite} for the recurrently coupled networks in our model. Each neuron $k$ is realized as a discrete-time non-leaky integrate and fire neuron with membrane potential $u_k\in [0,1)$. External input is given to the network in form of a (time-varying) Bernoulli process: At each time step, a single unit is randomly chosen and its membrane potential is increased by $u_0$. With a small time step $\Delta t$, this procedure approximates a Poisson process with intensities for each unit proportional to the probability that it will be chosen to receive external input. We also allow the situation in which certain units never receive external input, and that there is a positive probability that no unit receives external input at a time step.

If the external input pushes the membrane potential of the chosen unit above the threshold value $U_i$, an avalanche is started. At each generation of the avalanche, all units with membrane potentials above threshold fire a spike. The currently supra-threshold units are tracked by the spike activity vector $\VA(\Vu)\coloneqq \delta(\{i\in [N]:u_i \geq 1\})$. The membrane potentials of the spiking neurons are reset by subtracting $1$ and the membrane potentials of connected neurons are increased according to the network coupling matrix $W$ by $W\VA(\Vu)$. If after distribution of internal activation no unit is above threshold anymore, the avalanche stops. External input only occurs in-between of avalanches (separation of timescales). 

The exact mathematical definition of this random dynamical system and its formal treatment are given in~\citep{schunemann2022rigorous}.

%
%
%

\subsection{Model architecture}

The model consists of five subnetworks labeled $A,a,B,b,C$ of sizes $n,m,n,m,n$, respectively, which are recurrently coupled all-to-all within each subnetwork (homogeneous coupling). Connections between subnetworks have also homogeneous weights and are formed from each neuron in the sending network to each neuron in the receiving network. An isolated homogeneous network of size $n$ is at the critical point for the constant value $\alpha(n) = (1-1/\sqrt{n})/n$~\citep{eurich2002finite} for all entries of the coupling matrix. The coupling matrix $W$ of this model has the following block structure, where each entry represents the values of a constant matrix of the size determined by the subpopulations indicated on the labels on the rows and columns.
\begin{align}\label{eq:Wfull}
W = \bordermatrix{ & A & a & B & b & C \cr
A  & w_{A,A}\alpha(n)  & w_{A,a}    & 0                 & 0          & 0                \cr
a  & 0                 & \alpha(m)  & 0                 & 0          & 0                \cr
B  & 0                 & 0          & w_{B,B}\alpha(n)  & w_{B,b}    & 0                \cr
b  & 0                 & 0          & 0                 & \alpha(m)  & 0                \cr
C  & w_{C,A}\alpha(n)  & 0          & w_{C,B}\alpha(n)  & 0          & w_{C,C}\alpha(n) \cr
}
\end{align}
Default values for the coupling weights are listed in table~\ref{tab:model-params}.


\subsection{Input signals}

To model the situation in~\citep{grothe2018attention}, we drive the model by two input signals. The input signals consist of a mean drive modulated by independent input 'flicker' signals $f_A(t)$, $f_B(t)$. The two flicker signals are uncorrelated and assume independently drawn random values from a uniform distribution supported on $n_l$ equally spaced levels between $-1$ and $1$ every $10\unit{ms}$. Mean drive and flicker determine the input in each time step: the probabilities that a unit of subnetwork $A,\, B,\, a\, b$ receives external input at any given step are given by
\begin{align}
    p_A(t) &= n(1+cf_A(t))/p_0,\nonumber\\
    p_B(t) &= n(1+cf_B(t))/p_0,\nonumber\\
    p_a(t) &= \delta_a m(1+cf_A(t))/p_0,\nonumber\\ 
    p_b(t) &= \delta_b m(1+cf_B(t))/p_0,\nonumber 
\end{align}
respectively, with normalization $p_0 = 2(n+m)(1+c)+n$.  In particular, we realized this random drive such that in one time step either exactly one unit was driven with probability $p_A+p_B+p_a+p_b$, 
or that no unit was driven with probability $1-p_A+p_B+p_a+p_b$. Each unit in a subnetwork is equally likely to receive external input.

The requirement that external drive never activates two or more units simultaneously was necessary for allowing formal analysis. The choice of normalization opens the possibility to also include input drive to population $C$, but was never used in our simulations or analysis. $\delta_a$ and $\delta_b$ model the attentional condition and are explained in the next subsection.

\subsection{Attentional modulation}

The control populations $a$ or $b$ are normally inhibited and become only active if the corresponding input signal is attended. This means that the control population processing the non-attended stimulus is completely silenced, which is formally equivalent to the case where this population is not driven by external input. This equivalence alleviates the need to explicitly model inhibition. Instead we introduced the variables $\delta_a$ and $\delta_b$ for switching input to the control populations on or off. These variables take the value $1$ when attention is devoted to the corresponding input signal, and $0$ else.

\subsection{Synchronicity gain}

Pyramidal neurons in vivo respond much stronger to synchronous input compared to input of the same strength but with spike times distributed independently over a longer time interval. To model this effect, we introduce a threshold $\theta$ for the connections from V1 populations $A$ and $B$ to V4 population $C$. Only avalanches in $A$ or $B$ with a size of at least $\theta$ are passed through to $C$.
%

Note that for $\theta=1$, driving $V4$ with the total internal activation elicited from avalanches in $V1_A,V1_B$ results in a different model dynamics within one avalanche compared to the original EHE-Model without synchronicity gain mechanism, but it follows from~\citep[Corollary A.3]{schunemann2022rigorous} that both dynamics elicit identical distributions of avalanche assemblies. 

\subsection{Analytical avalanche size distributions}
\label{sec:theory-avs}


After specifying the coupling matrix and the input statistics of the model network, we can apply the mathematical framework developed in~\citep{schunemann2022rigorous} to compute the probability of avalanches in this model in closed form. W.l.o.g. we assume that attention is devoted to stimulus $A$, which is equivalent to control population $b$ being silent and not taking part in, or causing an avalanche.

Since our model consists of homogeneously connected subnetworks, we can simplify the resulting expressions by reducing determinants of sub-matrices of the full matrix $W$ to determinants of the $4\times 4$ matrix of block entries $\Wr$ obtained by selecting just one row/column per subnetwork $A,B,a,C$ from $W$ (see~\citep[Appendix section H.2]{schunemann2022rigorous}). 
In fact, $\Wr$ then looks like $W$ in \Eq{eq:Wfull} with the row/column for control population $b$ deleted, and its entries not representing full matrices but scalars.

The notation $\Wr_{|_{I}}$ forms a submatrix of $\Wr$ by selecting the set of rows and columns in a subset $I\subseteq \{A,B,a,C\}$ of the input-driven populations. 
We introduce the abbreviation $\mathcal{V}_k(\VN, \VU, W)$ for vectors $\VN = [N_1, N_2, \ldots N_k]$, and $\VU = [U_1, U_2, \ldots U_k]$ in $\mathbb{R}^k$ and a $k\times k$ matrix $W$:
\begin{align} \label{eq:vrbs}
    \mathcal{V}_k(\VN ,\VU, W) \coloneqq \mbox{det}\left( \diag(\VU)-W\VN \right)
    \prod_{i=1}^k U_{i}^{N_{i}-1}
\end{align}

We will now derive the avalanche size distribution of the subnetworks for the case $\theta=1$. Let us first condition on the event that a unit in the subnetwork $B$ starts the avalanche. The probability $P_B(n_B, n_C)$ that the resulting avalanche consists of $0 < n_B \leq n$ units in $B$ and $0 \leq n_C \leq n$ units in $C$ is given by
\begin{align}
    P_{B}(n_B, n_C) =& {1 \over P^{(s)}_B}  \binom{n}{n_B} \binom{n}{n_C}
        \mathcal{V}_2 \left( [n_B-1, n_C]^T, \Wr_{|_{B,C}} \, [n_B, n_C]^T,
        \Wr_{|_{B,C}} \right)\nonumber \\
    & \mathcal{V}_4 \left( [n-n_B, n, m, n-n_C]^T,
        \1-\Wr \, [n_B, 0, 0, n_C]^T, \Wr \right)
      \label{eq:PV1B}\\
    \mbox{with } P^{(s)}_B =& \mathcal{V}_4([n-1, n, m ,n]^T, \1, \Wr)\label{eq:PV1Bs} \text{ .}
\end{align}
    
Here, $P^{(s)}_B$ is the normalization constant of this distribution, i.e. the sum over all $P_B(n_B, n_C)$ for $0 < n_B \leq n$ and $0 \leq n_C \leq n$. Due to the symmetry in the coupling matrix, the probability of avalanches with $n_A$ units in $A$ and $n_C$ units in $C$, given that the avalanche started in $A$ is 
\begin{align}
    P_A(n_A, n_C) &= P_B(n_B, n_C) \label{eq:PV1A}\\
    P^{(s)}_A &= P^{(s)}_B\label{eq:PV1As} \text{ .}
\end{align}
If external input is given to a unit in the control population $a$, the resulting avalanche may consist of units in $a, A$ and $C$. The probability $P_{a}(m_a, n_A, n_C)$ of an avalanche consisting of $m_a$ units in $a$, $n_A$ units in $A$ and $n_C$ units in $C$ given, that the avalanche started in subnetwork $a$ is given by
\begin{align}
    P_{C}(m_A, n_A, n_C) =& {1 \over P^{(s)}_C} 
        \binom{m}{m_a}\binom{n}{n_A}\binom{n}{n_C}
        \mathcal{V}_3 \left( [n_A, m_a, n_C]^T, \Wr_{|_{A, a, C}}(n_A, m_a, n_C)^T,
        \Wr_{|_{A, a, C}} \right) \nonumber \\
    & 
        \mathcal{V}_4( [n, n-n_A, m-m_a, n-n_C]^T, \1-\Wr \, [0, n_A, m_a, n_C]^T, \Wr ) \label{eq:PC}\\
    \mbox{with } P^{(s)}_{C} =& \mathcal{V}_4( [n, n, m-1, n]^T, \1, \Wr )\label{eq:PCs} \text{ .}
\end{align}

\subsection{Avalanche size distributions for single populations}
\label{sec:theory-avs-singlepop}

Avalanche distribution for each individual population were derived from the joint probabilities $P^{\theta}_A(n_A, n_C), P^{\theta}_B(n_B, n_C), P^{\theta}_a(m_a, n_A, n_C)$ given that an avalanche is triggered by external input to population $A, B, a$, respectively, by the following marginalization equations:

$B$ is activated only by external input to $B$:
\begin{align}\label{eq:avsV1B}
    P^{\theta}_B(n_B) = \sum_{n_C=0}^{n} P^{\theta}_B(n_B, n_C) \text{ .} 
\end{align}

$a$ is activated only by external input to $a$:
\begin{align}\label{eq:avsC}
    P^{\theta}_{a}(m_a) = \sum_{n_A=0}^n\sum_{n_C=0}^n P^{\theta}_{a}(m_a, n_A, n_C)\text{ ,}
\end{align}

$A$ is activated either by external input to $A$ or to $a$:
The avalanche size distribution of the attended V1 population $A$ becomes a mixture of the corresponding marginal distributions conditioned on the event that at least one unit in $A$ participated in the avalanche,
\begin{align}\label{eq:avsV1A}
    P^{\theta}_A(n_A) =
        & \frac{m}{n+m} \sum_{m_a=1}^{m} \sum_{n_C=0}^n P^{\theta}_{a}(m_a, n_A, n_C)/
        \left( 1-\sum_{m_a=1}^{m} P^{\theta}_{a}(m_a, 0, 0) \right)+\nonumber \\
        & \frac{n}{n+m}\sum_{n_C=1}^n P^{\theta}_A(n_A, n_C) \, .
\end{align}

$C$ is activated either by external input to $A$, $B$ or to $a$:
Similarly to \Eq{eq:avsV1A}, the avalanche size distribution of the V4 population $C$ is a mixture of the corresponding marginals conditioned on $n_C>0$:
\begin{align}\label{eq:avsV4}
    P^{\theta}_{C}(n_C) =
    & \frac{n}{2n+m} \sum_{n_B=1}^{n} P^{\theta}_B(n_B, n_C)/
    \left( 1-\sum_{n_B=1}^n P^{\theta}_B(n_B, 0) \right)+ \nonumber\\
    & \frac{n}{2n+m} \sum_{n_A=1}^{n} P^{\theta}_A(n_A, n_C)/
    \left( 1-\sum_{n_A=1}^n P^{\theta}_A(n_A, 0) \right)+ \nonumber\\
    & \frac{m}{2n+m} \sum_{m_a=1}^{m} \sum_{n_A=1}^n P^{\theta}_{a}(m_a, n_A, n_C)/
    \left( 1-\sum_{m_a=1}^{m} \sum_{n_A=0}^n P^{\theta}_a(m_a, n_A, 0) \right) \, .
\end{align}

\subsection{Spectral and phase coherence}

\textbf{Spectral coherence.} Similar as in~\citep{grothe2018attention} we compute frequency-resolved temporal correlations using the wavelet-spectra of two signals. In~\citep{grothe2018attention}, one of these signals was the local field potential (LFP) measured in visual area V4. To obtain a proxy for the LFP from the mean firing rates in our network, we convolved the population firing rates with an exponential kernel $1/\tau \exp(-t/\tau)$ with time constant $\tau=15\unit{ms}$. We then computed a time-frequency resolved representation of a signal (LFP or visual stimulus) via a wavelet transformation using the Morlet wavelet with wave number $\omega=6$. Only values outside the cone of influence were used in computing the following measures. 

Given two wavelet-transformed signals $a_1(t, f)$ and $a_2(t, f)$, the spectral coherence $SC_{a_1, a_2}(f)$ between these signals is given by 
\begin{align}\label{eq:defSC}
    SC_{a_1, a_2}(f) = \frac{|\sum_t a_1(t, f)\overline{a_2(t, f)} |^2}
        {\left(\sum_t |a_1(t, f)|^2\right) \left( \sum_t |a_2(t, f)|^2\right) },
\end{align}
where $|\cdot|$ denotes the absolute value of a complex number and $\overline{\cdot}$ its complex conjugate.

\textbf{Phase coherence.} In addition to the spectral coherence, we also evaluated the phase coherence measure $pc(a_1, a_2, f)$ between two wavelet-transformed signals
$a_1$ and $a_2$ at frequency $f$ via
\begin{align}\label{eq:defPC}
    PC(a_1, a_2, f) = |1/N \sum_{t=1}^N \exp{i(\phi_{a_1(t, f)}-\phi_{a_2(t, f)})}| -           \frac{\sqrt{\pi}}{2\sqrt{N}}\text{ ,}
\end{align}
where $\phi$ represents the angle of the complex wavelet coefficients, and the second term a correction factor which leads to an unbiased estimator for finite $N$. The same method was used in~\citep{grothe2012switching}.

\subsection{Simulation parameters}

If not noted otherwise, we report results averaged over 15 trials of model simulations each of a duration of $T=250$ seconds. We bin the population activity in $A, B ,a ,C$ by computing the mean firing rate for each population binned to  $1\unit{ms}$ intervals and refer to the binned firing rates by $r_X(t)$. The same binning is applied to the flicker signals. \Tab{tab:model-params} summarizes the default parameter values of the model, which are used if not specified otherwise.

 We choose a sufficiently small time step of $\Delta t=1\mu s$ for a routing network with population sizes $n=100$ and $m=10$ in order to achieve a good approximation to a Poissonian input statistics. 
 The external input strength $u_0$ was chosen such that the average firing rate of V1 population $B$ was set to $r_B$, measured in Hz, using the analytical result from~\citep[equation~(12)]{schunemann2022rigorous}:
\begin{equation}
    u_0 = p_B r_B n (1-\alpha_c(n) w_{B,B} n) \Delta t \text{ .}
\end{equation}

 We evaluate the spectral coherence for the same pairs of signals as for the correlation coefficient on $80$ logarithmically spaced frequencies ranging from $\unit[5]{Hz}$ to $\unit[100]{Hz}$.

\begin{table}[htb]
\centering
\begin{tabular}{ l | l  }
Parameter description & default value \\
\hline
Simulation time step $\Delta t$ & $1\mu s$\\
Network sizes $n,m$ & $100,10$ \\
Recurrent coupling strength in V1 & $\beta=w_{A,A}=w_{B,B}=0.75$\\
Coupling from V1 to V4 & $w_{C,A}=w_{C,B}=0.3$\\
Coupling from control population to V1 & $w_{A,a}=w_{B,b}=0.095/m$\\
Recurrent coupling strength in V4 & $w_{C,C}=0.4$\\
Flicker modulation strength & $c=0.25$\\
Number of flicker levels & $n_l=5$\\
Firing rates $r_A,r_B$ (attended/non attended) & $\unit[40]{Hz},\unit[52]{Hz}$\\
Threshold for connections V1 to V4 connections & $\theta=5$

\end{tabular}
\caption{\label{tab:model-params}Table of default model parameter values}
\end{table}

\section*{Funding}
This work was supported by the Deutsche Forschungsgemeinschaft (DFG) via the Priority
Program 2205, Grant ER 324/5-1.

\bibliography{references}

\newpage

\section*{Figures}

\end{document}